\documentclass[a4paper,11pt]{article}
\usepackage{newlfont}
\usepackage{amssymb}
\usepackage{amsfonts}
\usepackage{amsmath}
\usepackage{amsthm}
\usepackage{youngtab}
\usepackage[all]{xy}

\newcommand{\tr}{\mbox{Tr}}
\newcommand{\bra}[1]{\ensuremath{\langle #1 |}}
\newcommand{\ket}[1]{\ensuremath{| #1 \rangle}}
\newcommand{\bk}[2]{\ensuremath{\langle #1 | #2 \rangle}}
\newcommand{\kb}[2]{\ensuremath{| #1 \rangle\!\langle #2 |}}

\newtheorem{theo}{Theorem}[section]

\newtheorem{cor}{Corollary}[section]

\theoremstyle{definition}
\newtheorem{example}{Example}[section]
\newtheorem{definition}{Definition}[section]
\newtheorem{remark}{Remark}[section]

\newcommand{\ot}{\otimes}
\newcommand{\ti}{\times}

\newcommand{\cL}{{\cal L}}
\newcommand{\cJ}{{\cal J}}

\newcommand{\cH}{{\cal H}}

\newcommand{\ra}{\rightarrow}

\newcommand{\C}{{\mathbb C}}

\newcommand{\wh}{\widehat}

\newcommand{\we}{\wedge}

\begin{document}

\title{Entanglement for multipartite systems\\ of indistinguishable particles
\thanks{Research of the first two authors financed by the Polish
Ministry of Science and Higher Education 
under the grant No. N N202 090239.} }
\author{Janusz Grabowski\footnote{email: jagrab@impan.pl} \\
\textit{Polish Academy of Sciences, Institute of Mathematics,} \\
\textit{\'Sniadeckich 8, P.O. Box 21, 00-956 Warsaw, Poland}          \\
\\
Marek Ku\'s\footnote{email: marek.kus@cft.edu.pl}\\
\textit{Center for Theoretical Physics, Polish Academy of Sciences,} \\
\textit{Aleja Lotnik{\'o}w 32/46, 02-668 Warszawa,
Poland} \\
\\
Giuseppe Marmo\footnote{email: marmo@na.infn.it}              \\
\textit{Dipartimento di Scienze Fisiche, Universit\`{a} ``Federico II'' di Napoli} \\
\textit{and Istituto Nazionale di Fisica Nucleare, Sezione di Napoli,} \\
\textit{Complesso Universitario di Monte Sant Angelo,} \\
\textit{Via Cintia, I-80126 Napoli, Italy} \\
}
\date{}
\maketitle

\begin{abstract}
We analyze the concept of entanglement for multipartite system
with bosonic and fermionic constituents and its generalization to
systems with arbitrary parastatistics. We use the representation
theory of symmetry groups to formulate a unified approach to this
problem in terms of simple tensors with appropriate symmetry. For
an arbitrary parastatistics, we define the S-rank generalizing the
notion of the Schmidt rank. The S-rank, defined for all types of
tensors, serves for distinguishing entanglement of pure states.
In addition, for Bose and Fermi statistics, we construct an
analog of the Jamio\l kowski isomorphism.



\end{abstract}

\section{Introduction}
Confronted with entangled states of a bipartite system,
Schr\"odinger said that entanglement was not
...\textit{\textbf{one} but rather \textbf{the} characteristic
trait of quantum mechanics, the one that enforces its entire
departure from classical line of thought}... \cite{schrodinger35}.

Note that Schr\"odinger's interests for this aspect of Quantum
Mechanics originated from the EPR paper. Indeed, the entanglement
for states of composite systems created quite an embarrassment for
theoretical physicists not sharing the point of view of the
``Copenhagen interpretation'' of quantum mechanics. As a matter of
fact, the entanglement played a rather important role in the
development of quantum mechanics, as it obliged the physics
community to face the nonlocal nature of the description of
natural processes. Nowadays, however, entanglement is considered
as an important resource for quantum computation, quantum
information and quantum teleportation.

The usual definition of entangled states is given by defining
first separable pure states as decomposable tensor products and
then declaring a state to be entangled if it is not separable.

In the case of a system containing identical constituents, the
problem of the definition of entanglement has to be reconsidered,
since the indistinguishability produces some amount of
entanglement by its own. For instance, a non-zero skew-symmetric
2-tensor is never decomposable. We cannot therefore simply
transpose to the Hilbert space of the composite system the
requirement that non-entanglement is guaranteed by the
factorization of the total state into a simple tensor product of
vectors corresponding to the two subsystems. We have to analyze
better the meaning of entanglement {\it per se} and refine the
concept of separability.

Several nonequivalent ways of identification and quantification of
the phenomenon were proposed. In \cite{herbut87} the authors
explicitly stated the problem of distinguishing non-local
correlations from those caused by the Pauli principle measured in
fundamental experiments, aiming at checking the validity of
quantum mechanics involving two identical fermions. To quantify
the amount of extra correlations, the authors used the number of
terms in the decomposition of the wave function in terms of
elementary $2\times 2$ `Slater determinants' (see
Section~\ref{sec:quadratic} below). Similar ideas, based on the
lengths of the canonical decomposition of the wave function, were
used in \cite{grobe94} for description of correlations in double
ionization of atoms in strong electromagnetic fields.

A scheme of extracting the correlations caused by the Pauli
principle in fermionic systems in order to characterize a genuine
entanglement, became important in analysis of elementary quantum
gates based on quantum dots \cite{schliemann01}. Subsystems here
are no longer separated by macroscopic distances; on the contrary,
they occupy the same spatial regions and their
indistinguishability becomes relevant. The problem was analyzed in
\cite{sckll01} and further in \cite{eckert02} in terms of modern
theory of entanglement. Measures of correlations were constructed
in analogy with the distinguishable particles case, again by
employing algebraic properties of the coefficient matrix in the
expansion of a state in terms of basis states.

Similar ideas can be applied, and in fact were applied, to bosons
\cite{eckert02,paskauskas01}. In the bosonic case non-entangled
states are again identified with simple symmetric tensors (see
Section~\ref{sec:c-rank} below). Thus, in the case of two bosons a
state is non-entangled if and only if it is a tensor product of
two identical one-particle states. This definition should be
contrasted with another one advocated already in \cite{herbut01},
where bosonic and fermionic systems were treated in parallel. To
identify two partite systems with \textit{minimal
identical-particle correlations}, i.e., without entanglement, one
considers a situation when two particles are confined to two
non-overlapping spatial regions in which they are subjected to
spatially confined independent measurements. The correlations are
minimal if, for every independent choice of measurements, the
probability of outcomes factorizes into the product of single
particles probabilities of outcomes. For two fermions, the result
reduces to the demand that the wave function of the whole system
is a simple antisymmetric tensor (\textit {a single Slater
determinant}),
$\ket{\Psi}\sim\left(\ket{\psi}\otimes\ket{\phi}-\ket{\phi}\otimes\ket{\psi}\right)$,
i.e., to the very definition adopted by other authors cited above.
For bosons, however, the definition leads to two different classes
of minimally correlated state. In addition to tensor products of
identical one-particle states, also wave functions in the form of
\textit{Slater permanents} i.e., symmetrized tensor products of
orthogonal one-particle states,
$\ket{\Psi}\sim\left(\ket{\psi}\otimes\ket{\phi}+
\ket{\phi}\otimes\ket{\psi}\right)$, $\bk{\psi}{\phi}=0$, are
classified as minimally entangled.\footnote{To make the
non-entanglement condition in bosonic and fermionic systems
completely parallel, one can demand orthogonality also in the
latter case, since it does not play a role for fermions.} The
definition can be easily extended to many particle systems
\cite{li01}.

In a series of papers by Ghirardi et. al.
\cite{ghirardi02,ghirardi04,ghirardi05}, a definition of
non-entangled state was based on the concept of a \textit{complete
set of properties} possessed by a constituent of a composite
system. For a bi-partite system, in a pure state $\rho$, we say
that a one of the constituents has a complete set of properties if
and only if there exists a rank-one projection operator $P$ acting
in a single particle space $\mathcal{H}$ such that $\tr(E\rho)=1$,
where $E=P\otimes I+I\otimes P-P\otimes P$ is a projection
operator acting in $\mathcal{H}\wedge\mathcal{H}$ or
$\mathcal{H}\vee\mathcal{H}$ for, respectively, fermions and
bosons. This idea resembles the one advocated in the above cited
papers of Herbut, but rather than invoking directly `local'
measurements it stresses properties of subsystem states. The
result is the same. For fermions, non-entangled states are
antisymmetrized tensor products, whereas for bosons they split
into two classes: products of two identical states, or symmetrized
products of two orthogonal states. Whereas for fermions the
criterion of non-entanglement, based on the number of coefficients
in expansion in terms of elementary $2\times 2$ Slater
determinants, remains functional, it is not the case for bosons.
Here, states expressible as linear combinations of two Slater
permanents are non-entangled if the both coefficients of expansion
are the same -- such states can be expressed as symmetrized
products of two orthogonal vectors \cite{li01,ghirardi02}. A concept of entanglement for indistinguishable particles based on a measurability of correlations was recently considered also in \cite{sasaki10}.

In this paper we attempt to approach this problem from a
mathematical view point, where structures and available
mathematical constructions are used as a guide. A fundamental
character of our work depends also on reviewing and generalizing
basic concepts of mathematical foundations in understanding
entanglement. We shall provide few physical considerations in the
conclusions, where we also briefly compare our results with other
outlined above. Here, let us mention only that our natural and
unifying mathematical model strongly suggests non-entanglement for
bosons to be associated with tensor products of identical states.

The paper is organized in the following way. In
Section~\ref{sec:tensoralgebras} we recall definitions and basic
facts from tensor algebra underlying our analysis of entanglement
for multipartite bosonic and fermionic systems. In
Section~\ref{sec:duality} we introduce the concept of duality and
analyze contractions between dual spaces of tensors, which allows
us to define the S-rank of a tensor, generalizing the Schmidt
rank, and then the simplicity of a tensor in
Section~\ref{sec:c-rank}. Section~\ref{section:characterizations}
contains various characterizations of simplicity of a tensor in
general, and in the bosonic or fermionic class. In
Section~\ref{sec:quadratic} we define entanglement for bosonic and
fermionic multipartite states and provide a simple
characterization of entanglement for pure states in terms of
bilinear functions in coefficients of representing tensors.
In analysis of entanglement of distinguishable particles, the so
called Jamio\l kowski isomorphism played the role of a very useful
tool. We give an extension of it for boson and fermions in
Section~\ref{sec:jamiol}. The mathematically rigorous
generalization of entanglement to multipartite system with
arbitrary parastatistics, another novel invention in this paper,
is given in Section \ref{sec:segrepara}.

Note that such an unifying approach allowed also for description of
pure non-entangled states as the images of generalized Segre maps,
in full analogy with the case of distinguishable particles \cite{grabowski05, grabowski06}.
These questions, however, we decided to postpone to a separate paper.

\section{Tensor algebras}
\label{sec:tensoralgebras}

To describe some properties of systems composed of
indistinguishable particles and to fix the notation, let us start
with introducing corresponding tensor algebras associated with a
Hilbert space $\mathcal{H}$. For simplicity, we assume that
$\mathcal{H}$ is finite-, say, $n$-dimensional, but a major part
of our work remains valid also for Hilbert spaces of infinite
dimensions. Note only that in the infinite dimensions the
corresponding tensor product $\mathcal{H}_1\otimes\mathcal{H}_2$
is the tensor product in the category of Hilbert spaces, i.e.,
corresponding to Hilbert-Schmidt norm.

In the tensor power $\mathcal{H}^{\otimes
k}=\underset{k-\mathrm{times}}{\underbrace{\mathcal{H}\otimes\cdots\otimes\mathcal{H}}}$,
we distinguish the subspaces: $\mathcal{H}^{\vee
k}=\underset{k-\mathrm{times}}{\underbrace{\mathcal{H}\vee\cdots\vee\mathcal{H}}}$
of totally symmetric tensors, and $\mathcal{H}^{\wedge
k}=\underset{k-\mathrm{times}}{\underbrace{\mathcal{H}\wedge\cdots\wedge\mathcal{H}}}$
of totally antisymmetric ones, together with the symmetrization,
$\pi_k^\vee:\mathcal{H}^{\otimes k}\rightarrow\mathcal{H}^{\vee
k}$, and antisymmetrization, $\pi_k^\wedge:\mathcal{H}^{\otimes
k}\rightarrow\mathcal{H}^{\wedge k}$, projectors:
\begin{eqnarray}\label{sproj}
\pi_k^\vee(f_1\otimes\cdots\otimes f_k)&=&
\frac{1}{k!}\sum_{\sigma\in S_k}f_{\sigma(1)}\otimes\cdots\otimes
f_{\sigma(k)},
\\
\pi_k^\wedge(f_1\otimes\cdots\otimes f_k)&=&
\frac{1}{k!}\sum_{\sigma\in S_k}(-1)^\sigma
f_{\sigma(1)}\otimes\cdots\otimes f_{\sigma(k)}.
\end{eqnarray}
Here, $S_k$ is the group of all permutations
$\sigma:\{1,\ldots,k\}\rightarrow\{1,\ldots,k\}$, and
$(-1)^\sigma$ denotes the sign of the permutation $\sigma$. Note
that with every permutation $\sigma\in S_k$ there is a canonically
associated  unitary operator $U_\sigma$ on $\mathcal{H}^{\otimes
k}$ defined by $$U_\sigma(f_1\otimes\cdots\otimes
f_k)=f_{\sigma(1)}\otimes\cdots\otimes f_{\sigma(k)}\,,$$ so that
the map $\sigma\mapsto U_\sigma$ is an injective unitary
representation of $S_k\rightarrow U(\mathcal{H}^{\otimes k})$. We
will write simply $\sigma$ instead of $U_\sigma$, if no
misunderstanding is possible. Symmetric and skew-symmetric tensors
are characterized in terms of this unitary action by $\sigma(v)=v$
and $\sigma(v)=(-1)^\sigma v$, respectively, for all $\sigma\in
S_k$

We put, by convention, $\mathcal{H}^{\otimes 0}=\mathcal{H}^{\vee
0}=\mathcal{H}^{\wedge 0}=\mathbb{C}$. It is well known that the
obvious structure of a unital graded associative algebra on the
graded space $\mathcal{H}^\otimes
=\mathop\otimes\limits_{k=0}^\infty\mathcal{H}^{\otimes k}$ (the
{\it tensor algebra}) induces canonical unital graded associative algebra structures
on the spaces $\mathcal{H}^\vee
=\mathop\oplus\limits_{k=0}^\infty\mathcal{H}^{\vee k}$ (called
the {\it bosonic Fock space}) and $\mathcal{H}^\wedge
=\mathop\oplus\limits_{k=0}^\infty\mathcal{H}^{\wedge k}$ (called
the {\it fermionic Fock space}) of symmetric and antisymmetric
tensors, with the multiplications \begin{equation}\label{vee}
v_1\vee v_2=\pi^\vee(v_1\otimes v_2),
\end{equation}
\begin{equation}\label{wedge}
w_1\wedge w_2=\pi^\wedge(w_1\otimes w_2).
\end{equation}
Here, of course,

\begin{equation}\label{sym}
\pi^\vee=\mathop\oplus\limits_{k=0}^\infty
\pi_k^\vee:\mathcal{H}^\otimes\rightarrow\mathcal{H}^\vee,
\end{equation}
and
\begin{equation}\label{asym}
\pi^\wedge=\mathop\oplus\limits_{k=0}^\infty
\pi_k^\wedge:\mathcal{H}^\otimes\rightarrow\mathcal{H}^\wedge,
\end{equation}
are the symmetrization and antisymmetrization projections. Note
that the multiplication in $\mathcal{H}^\vee$ is commutative,
$v_1\vee v_2= v_2\vee v_1$, and the multiplication in
$\mathcal{H}^\wedge$ is graded commutative, $w_1\wedge
w_2=(-1)^{k_1\cdot k_2} w_2\wedge w_1$, for
$w_i\in\mathcal{H}^{\wedge k_i}$.

Denote with $\cH^*$ the complex dual space of $\cH$. The Hermitian
product $\bk{\cdot}{\cdot}$ on $\cH$ induces a canonical bijection
between $\cH$ and $\cH^*$ which, in the Dirac's notation, reads
\begin{equation*}
\mathcal{H}\ni \left| x\right\rangle \mapsto \left\langle
x\right|\in\mathcal{H}^{\ast }.
\end{equation*}
We must stress that this is not an isomorphism of complex linear
spaces, since the above map is anti-linear. Note, however, that
the symmetric tensor algebra $\mathcal{H}^\vee$ can be canonically
identified with the algebra $Pol(\mathcal{H}^*)$ of polynomial
functions on the complex dual $\mathcal{H}^*$ of $\mathcal{H}$.
Indeed, any $f\in\mathcal{H}$ can be identified with a linear
function $x_f$ on $\mathcal{H}^*$ by means of the canonical
pairing $\langle\,
,\rangle:\mathcal{H}\times\mathcal{H}^*\rightarrow\mathbb{C}$
between the dual spaces, as $x_f(y)=\langle f, y\rangle$ (we must
distinguish this pairing from the Hermitian product on $\cH$).
This can be extended to an isomorphism of commutative algebras in
which $f_1\vee\cdots\vee f_k$ corresponds to the homogenous
polynomial $x_{f_1}\cdots x_{f_k}$. Similarly, one identifies
$\mathcal{H}^\wedge$ with the Grassmann algebra $Grass(\cH^*)$ of
polynomial (super)functions on $\mathcal{H}^*$. Here, however,
with $f\in\mathcal{H}$ we associate a linear function $\xi_f$ on
$\mathcal{H}^*$ regarded as and odd function:
$\xi_f\xi_{f^\prime}=-\xi_{f^\prime}\xi_f$. In the language of
supergeometry one speaks about the purely odd manifold
$\Pi\mathcal{H}^*$ obtained from the standard (purely even) linear
manifold $\mathcal{H}^*$ by changing the parity of linear
functions. In this sense, $\mathcal{H}^\wedge$ is the algebra of
holomorphic (super)functions on the complex supermanifold
$\mathcal{H}^*$.

If we fix a basis $e_1,\ldots,e_n$ in $\mathcal{H}$ and associate
with its elements even linear functions $x_1,\ldots,x_n$ on
$\mathcal{H}^*$, and odd linear functions $\xi_1,\ldots,\xi_n$ on
$\Pi\mathcal{H}^*$, then
$\mathcal{H}^\vee\simeq\mathbb{C}[x_1,\ldots,x_n]$, i.e.\
$\mathcal{H}^\vee$ becomes isomorphic with the algebra of complex
polynomials in $n$ commuting variables. Similarly,
$\mathcal{H}^\wedge\simeq\mathbb{C}[\xi_1,\ldots,\xi_n]$, i.e.,
$\mathcal{H}^\wedge$ is isomorphic with the algebra of complex
Grassmann polynomials in $n$ anticommuting variables. The
subspaces $\mathcal{H}^{\vee k}$ and $\mathcal{H}^{\wedge k}$
correspond to homogenous polynomials of degree $k$. It is
straightforward that homogeneous polynomials $x_1^{k_1}\cdots
x_n^{k_n}$, with $k_1+\cdots+k_n=k$, form a basis of
$\mathcal{H}^{\vee k}$, while homogeneous Grassmann polynomials
$\xi_{i_1}\wedge\cdots\wedge\xi_{i_k}$, with $1\le
i_1<i_2<\cdots<i_k\le n$, form a basis of $\mathcal{H}^{\wedge
k}$. In consequence, $\dim\mathcal{H}^{\vee
k}={{n+k-1}\choose{k}}$ and $\dim\mathcal{H}^{\wedge
k}={{n}\choose{k}}$, so the gradation in the fermionic Fock space
is finite (for finite-dimensional $\mathcal{H})$. Of
course, we can put together both algebras and consider the tensor
product $\mathcal{H}_1^\vee\otimes\mathcal{H}_2^\wedge$, with
$\dim{H}_1=n$ and $\dim{H}_2=m$, which is a graded associative
algebra with a bi-gradation $\mathbb{N}\times\mathbb{N}$
concentrated in $\mathbb{N}\times\{0,1,\ldots,m\}$,
\begin{equation}\label{fulltensor}
\mathcal{H}_1^\vee\otimes\mathcal{H}_2^\wedge=\mathop\bigoplus\limits_{(k,l)\in\mathbb{N}
\times\{0,1,\ldots,m\}}\mathcal{H}_1^{\vee
k}\otimes\mathcal{H}_2^{\wedge l}.
\end{equation}
Here, $\mathcal{H}_1^{\vee k}\otimes\mathcal{H}_2^{\wedge l}$
represent systems composed of $k$ bosons, described by a Hilbert
space of dimension $n$, and $l$ fermions in the Hilbert space of
dimension $m$. The whole algebra
$\mathcal{H}_1^\vee\otimes\mathcal{H}_2^\wedge$ is the algebra of
polynomial functions on the linear supermanifold
$\mathcal{H}_1^*\times\Pi\mathcal{H}_2^*$ of dimension $(n,m)$.
Such functions are written as finite complex linear combinations
\begin{equation}\label{all}
\sum_{\substack{k_1,\ldots,k_n \\1\le i_1<\cdots<i_l\le
m}}a_{k_1,\ldots,k_n}^{i_1,\ldots,i_l} x_1^{k_1}\cdots
x_n^{k_n}\xi_{i_1}\cdots\xi_{i_l}.
\end{equation}

\medskip
Note that any basis $\{e_1,\ldots, e_n\}$ in $\mathcal{H}$ induces
a basis $\big\{e_{i_1}\otimes e_{i_2}\otimes\cdots\otimes
e_{i_k}\, |\; i_1,\ldots, i_k\in\{1,\ldots,n\}\big\}$ in
$\mathcal{H}^{\otimes k}$. Therefore, any
$u\in\mathcal{H}^{\otimes k}$ can be uniquely written as a linear
combination
\begin{equation}\label{gentens}
u=\sum_{i_1,\ldots,i_k=1}^n u^{i_1\ldots i_k}e_{i_1}
\otimes\ldots\otimes e_{i_k}.
\end{equation}
If $u\in\mathcal{H}^{\vee k}$, then the tensor coefficients
$u^{i_1\ldots i_k}$ are totally symmetric and, after applying the
symmetrization projection to (\ref{gentens}), we get
\begin{equation}\label{symtens}
u=\sum_{i_1,\ldots,i_k=1}^n u^{i_1\ldots i_k}e_{i_1}
\vee\ldots\vee e_{i_k}.
\end{equation}
Similarly, if $u\in\mathcal{H}^{\wedge k}$, the tensor
coefficients $u^{i_1\ldots i_k}$ are totally antisymmetric and
\begin{equation}\label{antisymtens}
u=\sum_{i_1,\ldots,i_k=1}^n u^{i_1\ldots i_k}e_{i_1}
\wedge\ldots\wedge e_{i_k}.
\end{equation}
We will refer to the coefficients $u^{i_1\ldots i_k}$ as to the
{\it coefficients of $u$ in the basis} $\{e_1,\ldots, e_n\}$.

\section{Tensor duality and contractions}
\label{sec:duality}

Starting with the canonical duality (pairing) $\langle\, ,\rangle$
between vectors from $\mathcal{H}$ and covectors from
$\mathcal{H}^*$, with an obvious prolongation to a paring between
$\mathcal{H}^{\otimes k}$ and $(\mathcal{H}^*)^{\otimes k}$,
\begin{equation}\label{kpairing}
\langle f_1\otimes\cdots\otimes f_k,  g_1\otimes\cdots\otimes
g_k\rangle =\prod_{i=1}^k\langle f_i, g_i\rangle,
\end{equation}
and viewing symmetric and antisymmetric tensors as canonically
embedded in the tensor algebra, we find canonical pairings between
$\mathcal{H}^{\vee k}$ and $(\mathcal{H}^*)^{\vee k}$, as well as
between $\mathcal{H}^{\wedge k}$ and $(\mathcal{H}^*)^{\wedge k}$.
For $f_1,\dots,f_k\in\cH$ and $g_1,\dots,g_k\in\cH^*$, we get
\begin{equation}\label{pairingvee}
\langle f_1\vee\cdots\vee f_k,  g_1\vee\cdots\vee g_k\rangle
=\frac{1}{(k!)^2}\sum_{\sigma,\tau\in S_k}\prod_{i=1}^k \langle
f_{\sigma(i)}, g_{\tau(i)}\rangle =\frac{1}{k!}\mathrm{per}
(\langle f_i, g_j\rangle).
\end{equation}
Here, $\sum_{\tau\in S_k}\prod_{i=1}^k
a_{i\tau(i)}=\mathrm{per}(a_{ij})$ is the permanent of the matrix
$A=(a_{ij})$. Similarly,
\begin{equation}\label{pairingwedge}
\langle f_1\wedge\cdots\wedge f_k,  g_1\wedge\cdots\wedge
g_k\rangle =\frac{1}{k!}\det(\langle f_i, g_j\rangle).
\end{equation}

Quite similarly one can prove that the natural Hermitian product
on $\mathcal{H}^\otimes$, for which the grading is the
decomposition into orthogonal subspaces and
\begin{equation}\label{paringtensor}
\bk{f_1\otimes\cdots\otimes f_k}{f_1^\prime\otimes\cdots\otimes
f_k^\prime} =\prod_{i=1}^k\bk{f_i}{f_i^\prime},
\end{equation}
induces Hermitian products on the subspaces $\mathcal{H}^{\vee k}$
and $\mathcal{H}^{\wedge k}$ which read, respectively,
\begin{eqnarray}
\bk{f_1\vee\cdots\vee f_k}{f_1^\prime\vee\cdots\vee f_k^\prime}
&=&\frac{1}{k!}\mathrm{per}(\bk{f_i}{f_j^\prime}), \\
\bk{f_1\wedge\cdots\wedge f_k}{f_1^\prime\wedge\cdots\wedge
f_k^\prime} &=&\frac{1}{k!}\det(\bk{f_i}{f_j^\prime}).
\end{eqnarray}

Given a basis $f_1,\ldots, f_n$ of $\mathcal{H}$ and the dual
basis $f_1^*,\ldots, f_n^*$ of $\mathcal{H}^*$, we have the
induced bases:
$$f_1^{k_1}\vee\cdots\vee f_n^{k_n}\,,\quad k_1+\cdots
+k_n=k$$ of $\mathcal{H}^{\vee k}$, and
$$\,f_{i_1}\wedge\cdots\wedge
f_{i_k}\,,\quad 1\le i_1<i_2<\cdots<i_k\le n$$ of
$\mathcal{H}^{\wedge k}$. The dual bases read
\begin{equation}\label{dualbvee}
\frac{k!}{k_1!\cdots k_n!}
\left(f_1^*\right)^{k_1}\vee\cdots\vee\left(f_n^*\right)^{k_n},
\quad k_1+\cdots +k_n=k,
\end{equation}
and
\begin{equation}\label{dualbvedge}
k!f_{i_1}^*\wedge\cdots\wedge f_{i_k}^*, \quad 1\le i_1<i_2<\cdots<i_k\le n.
\end{equation}

Analogously, any orthonormal basis $e_1,\ldots,e_n$ of
$\mathcal{H}$ induces canonical orthonormal bases
\begin{equation}\label{ortbvee}
\sqrt{\frac{k!}{k_1!\cdots k_n!}}\; e_1^{k_1}\vee\cdots\vee
e_n^{k_n}, \quad k_1+\cdots +k_n=k,
\end{equation}
of $\mathcal{H}^{\vee k}$, and
\begin{equation}\label{ortbwedge}
\sqrt{k!}\,e_{i_1}\wedge\cdots\wedge e_{i_k}, \quad 1\le
i_1<i_2<\cdots<i_k\le n\,,
\end{equation}
of $\mathcal{H}^{\wedge k}$.

The canonical pairings between $\mathcal{H}^{\vee k}$ and
$(\mathcal{H}^*)^{\vee k}$ on one hand, and $\mathcal{H}^{\wedge
k}$ and $(\mathcal{H}^*)^{\wedge k}$ on the other, can be
generalized to certain `pairings' ({\it contractions} or {\it
inner products}) between $\mathcal{H}^{\vee k}$ and
$(\mathcal{H}^*)^{\vee l}$ on one hand, and $\mathcal{H}^{\wedge
k}$ and $(\mathcal{H}^*)^{\wedge l}$ on the other. For the
standard simple tensors $f=f_1\otimes\cdots\otimes f_k\in
\mathcal{H}^{\otimes k}$ and $ g= g_1\otimes\cdots\otimes g_l\in
(\mathcal{H}^*)^{\otimes l}$, we just put
$$\imath_gf=\langle
f_1\otimes\cdots\otimes f_l,  g_1\otimes\cdots\otimes g_l\rangle
f_{l+1}\otimes\cdots\otimes f_k$$
if $l\le k$ and $\iota_gf=0$ if $l>k$, and extend it by linearity to
all tensors. It is easy to see now that, if $v=f_1\vee\cdots\vee
f_k\in \mathcal{H}^{\vee k}\subset\mathcal{H}^{\otimes k}$ and
$\nu=g_1\vee\cdots\vee g_l\in (\mathcal{H}^*)^{\vee l}
\subset(\mathcal{H}^*)^{\otimes l}$ then
$\imath_{\nu}v\in\mathcal{H}^{\vee (k-l)}$.

Similarly, $\imath_{\omega}w\in\mathcal{H}^{\wedge (k-l)}$, if
$w\in \mathcal{H}^{\wedge k}\subset\mathcal{H}^{\otimes k}$ and
$\omega\in (\mathcal{H}^*)^{\wedge l}
\subset(\mathcal{H}^*)^{\otimes l}$. Explicitly,
\begin{eqnarray}
\imath_{g_1\vee\cdots\vee g_l}f_1\vee\cdots\vee f_k
=\frac{1}{k!\,l!} \sum_{\substack{\sigma\in S_k\\ \tau\in
S_l}}\prod_{j=1}^l \langle f_{\sigma(j)},g_{\tau(j)}\rangle
f_{\sigma(l+1)}
\otimes\cdots\otimes f_{\sigma(k)} \nonumber \\
=\frac{(k-l)!}{k!} \sum_{\substack{S\in S(l,k-l)\\ \tau\in
S_l}}\prod_{j=1}^l \langle f_{S(j)},g_{\tau(j)}\rangle f_{S(l+1)}
\vee\cdots\vee f_{S(k)},
\end{eqnarray}
where $S(l,k-l)$ denotes the set of all $(l,k-l)$ shuffles.

For skew-symmetric tensors,
\begin{eqnarray}
\imath_{g_1\wedge\cdots\wedge g_l}f_1\wedge\cdots\wedge f_k
=\frac{1}{k!\,l!} \sum_{\substack{\sigma\in S_k\\ \tau\in
S_l}}(-1)^\sigma(-1)^\tau \prod_{j=1}^l\langle
f_{\sigma(j)},g_{\tau(j)}\rangle f_{\sigma(l+1)}
\otimes\cdots\otimes f_{\sigma(k)} \nonumber \\
=\frac{(k-l)!}{k!} \sum_{\substack{S\in S(l,k-l)\\ \tau\in
S_l}}(-1)^\sigma (-1)^\tau \prod_{j=1}^l\langle
f_{S(j)},g_{\tau(j)}\rangle f_{S(l+1)} \wedge\cdots\wedge
f_{S(k)}.
\end{eqnarray}
In particular,
\begin{equation}
\imath_{g_1\vee\cdots\vee g_k}f_1\vee\cdots\vee f_k = \langle
f_1\vee\cdots\vee f_k, g_1\vee\cdots\vee g_k \rangle=\frac{1}{k!}\mathrm{per}(\bk{f_i}{g_j})\,,
\end{equation}
and
\begin{equation}
\imath_{g_1\wedge\cdots\wedge g_k}f_1\wedge\cdots\wedge f_k =
\langle f_1\wedge\cdots\wedge f_k, g_1\wedge\cdots\wedge g_k
\rangle=\frac{1}{k!}\det(\bk{f_i}{g_j})\,.
\end{equation}
Moreover,
\begin{eqnarray}\label{c1}
\imath_{g_1\vee\cdots\vee g_{k-1}}f_1\vee\cdots\vee f_k &=&
\frac{1}{k}\sum_{j=1}^k \langle
f_1\vee\overset{\underset{\vee}{j}}{\cdots}\vee f_k,
g_1\vee\cdots\vee g_{k-1} \rangle f_j\\&=&
\frac{1}{k!}\sum_{j=1}^k\mathrm{per}(\bk{f_i}{g_s}_{i\ne j})f_j\,,\nonumber
\end{eqnarray}
and
\begin{eqnarray}\label{c2}
\imath_{g_1\wedge\cdots\wedge g_{k-1}}f_1\wedge\cdots\wedge f_k &=&
\frac{1}{k}\sum_{j=1}^k(-1)^{k-j} \langle
f_1\wedge\overset{\underset{\vee}{j}}{\cdots}\wedge f_k,
g_1\wedge\cdots\wedge g_{k-1} \rangle f_j\\&=&
\frac{1}{k!}\sum_{j=1}^k(-1)^{k-j}\det(\bk{f_i}{g_s}_{i\ne j})f_j\,.\nonumber
\end{eqnarray}

\section{The S-rank}
\label{sec:c-rank}

There are many concepts of a rank of a tensor used to describe its
complexity. One of the simplest and most natural is the one based
on the inner product operators defined in the previous section. We
will call it the \textit{S-rank}, since it turns out to be a
natural generalization of the {\it Schmidt rank} of 2-tensors.
\begin{definition}\label{def:c-rank}
Let $u\in\mathcal{H}^{\otimes k}$. By the {\it S-rank} of $u$, we
understand the maximum of dimensions
of the linear spaces $\imath_{\mathcal{H}}^{k-1}\sigma(u)$, for
$\sigma\in S_k$, which are the images of the contraction maps
\begin{equation}\label{1}
(\mathcal{H}^{\ast})^{\otimes (k-1)}\ni \mu\mapsto
\imath_{\mu}\sigma(u)\in\mathcal{H}.
\end{equation}
\end{definition}

\begin{theo}\label{theo:c-rank}
If $u\in\mathcal{H}^{\vee k}$ (resp., $u\in\mathcal{H}^{\wedge
k}$), then the S-rank of $u$ equals the dimension of
the linear space being the image of the contraction map
\begin{equation}\label{2}
(\mathcal{H}^{\ast})^{\vee (k-1)}\ni \mu\mapsto
\imath_{\mu}u\in\mathcal{H},
\end{equation}
(resp.,
\begin{equation}\label{3}
(\mathcal{H}^{\ast})^{\wedge(k-1)}\ni \mu\mapsto
\imath_{\mu}u\in\mathcal{H}).
\end{equation}
\end{theo}

\textbf{Proof:} It follows immediately from the observation that a
contraction of a symmetric (res., antisymmetric) tensor $u$ with a
tensor $\mu\in(\mathcal{H}^{\ast})^{\otimes i}$ is the same as its
contraction with the symmetrization (resp., antisymmetrization) of
$\mu$, $\imath_{\mu} u= \imath_{\pi^\vee (\mu)}u$, (resp.
$\imath_{\mu}u= \imath_{\pi^\wedge (\mu)}u$) and that, for
$\sigma\in S_k$, $\sigma(u)=\pm u$.

$\Box$

\begin{example}\label{we1} Let $e_1,e_2$ be orthogonal vectors in $\cH$. The vector
\begin{equation}\label{u0} u=e_1\vee e_2=\frac{1}{2}\left(e_1\otimes e_2+e_2\otimes e_1\right)
\end{equation}
has S-rank 2. Indeed, for each $x\in\cH$,
$$\imath_xu=\frac{1}{2}\imath_x\left(e_1\otimes e_2+e_2\otimes e_1\right)=\frac{1}{2}\left(\bk{x}{e_1}e_2+\bk{x}{e_2}e_1\right)\,,$$
so that $\imath_\cH u$ is spanned by $\{ e_1,e_2\}$.
\end{example}
\begin{example}\label{we2} Let us take $\cH$ of dimension 3 and an orthonormal basis $e_i$, $i=1,2,3$.
Let $u\in\cH^{\vee 4}$ in the polynomial notation reads
\begin{eqnarray}\label{u}
u&=&e_1^4+e_2^4+16e_3^4+4e_1^3e_2+8e_1^3e_3+4e_1e_2^3+8e_2^3e_3+32e_1e_3^3+32e_2e_3^3\\
&&+6e_1^2e_2^2+24e_1^2e_3^2+24e_2^2e_3^2+24e_1^2e_2e_3+24e_1e_2^2e_3+48e_1e_2e_3^2\,.   \nonumber
\end{eqnarray}
The vectors
$$e_1^3\,,e_2^3\,,e_3^3\,,e_1^2e_2\,,e_1^2e_3\,,e_1e_2^2\,,e_2^2e_3\,,e_1e_3^2\,,e_2e_3^2\,,e_1e_2e_3$$
form a basis of $\cH^{\vee 3}$ and it can be directly checked with the use of (\ref{c1}) that any of these vectors, say $\mu$, contracted with $u$ gives a vector proportional to $e_1+e_2+2e_3$. For instance, the only non-zero parts of $\imath_\mu u$ for $\mu=e_2^3$
are
$$\imath_{e_2^3}(u)=\imath_{e_2^3}\left(e_2^4+4e_1e_2^3+8e_2^3e_3\right)=e_2+e_1+2e_3$$
and
$$\imath_{e_1^2e_3}(u)=\imath_{e_1^2e_3}\left(8e_1^3e_3+24e_1^2e_3^2+24e_1^2e_2e_3\right)=
\Vert e_1^2e_3\Vert^2\left(6e_1+12e_3+6e_2\right)=2(e_1+2e_3+e_2)\,.
$$
It follows that $\imath_{\cH^{\vee 3}}(u)$ is 1-dimensional, so the S-rank of $u$ is 1.
\end{example}
\begin{example}\label{we3} For $\cH$ with the basis as above, consider $w\in\cH^{\wedge 2}$ of the form
\begin{equation}\label{w} w=e_1\we e_2+e_2\we e_3\,.
\end{equation}
Let us see that the S-rank of $w$ is 2. Indeed,
$$\imath_{e_1}w=\frac{1}{2}e_2\,,\quad \imath_{e_2}w=\frac{1}{2}\left(e_3-e_1\right)\,,\quad \imath_{e_3}w=-\frac{1}{2}e_2\,,$$
so that $\imath_\cH w$ is spanned by $e_2$ and $e_3-e_1$.
\end{example}
\begin{example}\label{we4}
For $\cH$ with an orthonormal basis $e_1,e_2,e_3,e_4$ the tensor
\begin{equation}\label{w1} w=e_1\we e_2+e_3\we e_4
\end{equation}
has the S-rank 4. Indeed,
$$\imath_{e_1}w=\frac{1}{2}e_2\,,\quad \imath_{e_2}w=-\frac{1}{2}e_1\,,\quad \imath_{e_3}w\frac{1}{2}e_4\,,\quad \imath_{e_4}w=-\frac{1}{2}e_3\,.$$
\end{example}

\begin{theo}\label{theo:simple}
\begin{description}
\item{(a)} The minimal possible S-rank of a non-zero tensor
    $u\in\mathcal{H}^{\otimes k}$ equals $1$. A tensor
    $u\in\mathcal{H}^{\otimes k}$ is of S-rank $1$ if and only if $u$
    is decomposable, i.e., \ $u$ can be written in the form
\begin{equation}\label{rank1}
u=f_1\otimes\cdots\otimes f_k, \quad f_i\in \mathcal{H},\quad
f_i\ne 0.
\end{equation}
Such tensors span $\mathcal{H}^{\otimes k}$. \item{(b)} The
minimal possible S-rank of a non-zero tensor
    $v\in\mathcal{H}^{\vee k}$ equals $1$. A tensor
    $v\in\mathcal{H}^{\vee k}$ is of S-rank $1$ if and only if $v$ can
    be written in the form
\begin{equation}\label{rank1a}
v=f\vee\cdots\vee f, \quad f\in \mathcal{H},\quad f\ne 0.
\end{equation}
Such tensors span $\mathcal{H}^{\vee k}$. \item{(c)} The minimal
possible S-rank of a non-zero tensor
    $w\in\mathcal{H}^{\wedge k}$ equals $k$. A tensor
    $w\in\mathcal{H}^{\wedge k}$ is of S-rank $k$ if and only if $w$
    can be written in the form
\begin{equation}\label{rank1b}
w=f_1\wedge\cdots\wedge f_k,
\end{equation}
where $f_1,\ldots,f_k\in\mathcal{H}$ are linearly independent.
Such tensors span $\mathcal{H}^{\wedge k}$.
\end{description}
\end{theo}

\textbf{Proof:} The `if' parts of the above statements are
obvious, so we shall prove `only if'. If the dimension of
$\imath_{\mathcal{H}}^{k-1}u$ is 1, thus the space is spanned by
some $f=f_k\in\cH$, then clearly $u=u'\otimes f$ for some
$u'\in\cH^{\otimes(k-1)}$. If, in turn, the tensor is symmetric,
then clearly $u$ is proportional to $f\otimes\cdots\otimes f$. In
the general case we get a similar fact for $\sigma(u)$ with
$\sigma\in S_k$, so $u=f_1\otimes\cdots\otimes f_k$. If $u$ is
skew-symmetric and $f_1,\dots,f_r$ span
$\imath_{\mathcal{H}}^{k-1}u$, then $u$ is a linear combination of
tensor products of these vectors, so of $f_{i_1}\wedge\cdots\wedge
f_{i_k}$. Hence, $r\ge k$, and $r=k$ if and only if $u$ is
proportional to $f_1\wedge\cdots\wedge f_k$. That simple tensors
span the corresponding spaces is pretty well known for general and
antisymmetric tensors. For symmetric tensors it follows from the
fact that powers of linear functions span the spaces of
polynomials with coefficients in a field of characteristic 0.

$\Box$

\begin{definition}\label{def:simple}
Tensors of minimal S-rank in $\mathcal{H}^{\otimes k}$ (resp.,
$\mathcal{H}^{\vee k}$, $\mathcal{H}^{\wedge k}$) we will call
{\it simple} (resp., {\it simple symmetric, simple
antisymmetric}).
\end{definition}
Theorem \ref{theo:simple} immediately implies the following.
\begin{cor} The S-rank is 1 for simple and simple
symmetric tensors, and it is $k$ for simple antisymmetric tensors
from $\cH^{\we k}$. Simple tensor have the form (\ref{rank1}),
simple symmetric tensors have the form (\ref{rank1a}), and simple
antisymmetric tensor have the form (\ref{rank1b}).
\end{cor}

\begin{example} The symmetric tensor $u$ defined in (\ref{u}), Example \ref{we2}, has the S-rank 1, so it is simple. As a matter of fact,
$$u=\left(e_1+e_2+2e_3\right)^4\,.$$
Also the antisymmetric tensor $w$ defined in (\ref{w}), Example \ref{we3}, is simple:
$$w=e_2\wedge(e_3-e_1)\,.$$
\end{example}

\begin{remark}\label{re:dist}
Of course, for distinguishable particles there is no need to use
the same Hilbert space $\mathcal{H}$ in the tensor products. We
can use the tensor product
$\mathcal{H}_1\otimes\cdots\otimes\mathcal{H}_k$ instead, with
simple tensors being decomposable: $u=f_1\otimes\cdots\otimes
f_k$, $f_i\in\mathcal{H}_i$.
\end{remark}

\section{Various characterizations of simple tensors}
\label{section:characterizations}

 For $\mathcal{H}_0^{\otimes 2}$, where
$\mathcal{H}_0=\mathcal{H}_1\otimes\cdots\otimes\mathcal{H}_k$, we
denote with $\sigma_i:\mathcal{H}_0^{\otimes 2}\ra
\mathcal{H}_0^{\otimes 2}$, $i=1,\ldots,k$, the transposition with
respect to the $i$-th and the $(k+i)$-th factor,
\begin{eqnarray}\label{lambda} &&\sigma_i(a_1\otimes\cdots\otimes
a_{2k})=
\\
&&a_1\otimes\cdots\otimes a_{i-1}\otimes a_{k+i}\otimes a_{i+1}
\otimes\cdots\otimes a_{k+i-1}\otimes a_i\otimes
a_{k+i+1}\otimes\cdots\otimes a_{2k}\,, \nonumber
\end{eqnarray}
and with $\tau_k:\mathcal{H}_0^{\otimes 2}\ra
\mathcal{H}_0^{\otimes 2}$ -- the cyclic permutation that moves
the last factor into the first place: $$
\tau_k(a_1\otimes\cdots\otimes a_{2k})= a_{2k}\otimes
a_1\otimes\cdots\otimes a_{2k-1}\,. $$

\begin{theo}\label{theo:simple1}
For $u\in\mathcal{H}_1\otimes\cdots\otimes\mathcal{H}_k$ the
following are equivalent:
\begin{description}
\item{(a)} $u$ is simple; \item{(b)} $\forall \sigma\in S_k\
\forall \ \mu_1,\mu_2\in
\mathcal{H}^*_{\sigma(1)}\otimes\cdots\otimes\mathcal{H}^*_{\sigma(k-1)}\quad
    \imath_{\mu_1}\sigma(u)\wedge\imath_{\mu_2}\sigma(u)=0$,
\item{(c)} $\sigma_i(u\otimes u)=u\otimes u$ for $i=1,\ldots,k$.
\end{description}
\end{theo}
\textbf{Proof:} (a) $\Rightarrow$ (b) is obvious in view of
Theorem~\ref{theo:simple} (a). Also (b) $\Rightarrow$ (c) is
clear, so assume $\sigma_k(u\otimes u)=u\otimes u$. This implies
that the dimension of the space $\imath_{\mathcal{H}}^{k-1}u$ is
1. Indeed, if this space is spanned by linearly independent
vectors $g_1,\dots,g_r$, then $u=\sum_{j=1}^ru_r\otimes g_r$ for
some linearly independent $u_j\in\cH^{\otimes(k-1)}$. Since
$\sigma_k(u\otimes u)=u\otimes u$ means that
$$\sum_{j,s=1}^ru_j\otimes g_j\otimes u_s\otimes g_s=
\sum_{j,s=1}^ru_j\otimes g_s\otimes u_s\otimes g_j,$$ we conclude
that $r=1$, so $u=u'\otimes f_k$ for some $f_k\in\cH$ and
$u'\in\cH^{\otimes(k-1)}$. A similar reasoning, applied to the
identities $\sigma_i(u\otimes u)=u\otimes u$ with $i=1,\dots,k-1$,
implies that $u=f_1\otimes\cdots\otimes f_k$.

$\Box$

\begin{theo}\label{theo:simplesym}
For a symmetric tensor $v\in\mathcal{H}^{\vee k}$ the following
are equivalent
\begin{description}
\item{(a)} $v$ is simple symmetric, \item{(b)} $\forall \
\nu_1,\nu_2\in (\mathcal{H}^*)^{\vee (k-1)}$
    $\imath_{\nu_1}v\wedge\imath_{\nu_2}v=0$,
\item{(c)} $v\otimes v=\sigma_k(v\otimes v)$.
\end{description}
\end{theo}
\textbf{Proof:} (a) $\Rightarrow$ (b) is obvious in view of
Theorem~\ref{theo:simple} (b). As (b) implies $v=v'\otimes f_k$
for some $f_k\in\cH$, also (b) $\Rightarrow$ (c) is clear. The
condition (c), in turn, for symmetric tensor yields $v\otimes
v=\sigma_i(v\otimes v)$ for all $i=1,\dots,k$, so
$v=f_1\otimes\cdots\otimes f_k$ as above, thus
$v=f\otimes\cdots\otimes f$ by symmetry.

$\Box$

\begin{theo}\label{theo:simpleantisym}
For an antisymmetric tensor $w\in\mathcal{H}^{\wedge k}$ the
following are equivalent
\begin{description}
\item{(a)} $w$ is simple antisymmetric, \item{(b)} $\forall \
\omega\in (\mathcal{H}^*)^{\wedge (k-1)}$
    $w\wedge\imath_{\omega}w=0$,
\item{(c)} $(\pi_{k+1}^\wedge\otimes
id_{\cH^{\otimes(k-1)}})(\tau_k(w\otimes w))=0$.
\end{description}
\end{theo}
\textbf{Proof:} (a) $\Rightarrow$ (b) is obvious in view of
Theorem~\ref{theo:simple} (c). Also (b) $\Rightarrow$ (a) is clear
and well known. We shall show that (b) and (c) are equivalent.

Let us write $w$ as a sum of simple antisymmetric tensors,
$w=\sum_{j=1}^mw_j$, with minimal $m$. Write
$w_j=f_j^1\wedge\cdots\wedge f_j^k$ and denote
$w_j^s=f_j^1\wedge\cdots \wh{f_j^s}\wedge\cdots\wedge f_j^k$,
$s=1,\dots,k$. Here, "\ {$\wh{}$}\ " stands for the omission. As
the number of simple tensors is minimal, the tensors $w_j^s$ are
linearly independent in $\cH^{\wedge(k-1)}$.

Observe now that
$$w_j=\sum_s\frac{(-1)^{k-s}}{k}w_j^s\otimes f_j^s\,,$$
so that
$$(\pi_{k+1}^\wedge\otimes id_{\cH^{\otimes(k-1)}})(\tau_k(w\otimes w))=
\sum_{j,s}\frac{(-1)^{k-s}}{k}f_j^s\wedge w\otimes w_j^s\,.$$
Since $w_j^s$ are linearly independent, the latter vanishes if and
only if all $f_j^s\wedge w$ vanish, that is clearly equivalent to
(b).

$\Box$

\begin{remark} The conditions (c) in Theorems \ref{theo:simple1},
\ref{theo:simplesym}, and \ref{theo:simpleantisym} have formally
this advantage over the corresponding conditions (b) that they are
directly verifiable, as they do not contain general quantifiers
referring to infinite sets.
\end{remark}

Note that all we have said remains valid for an arbitrary vector
space over a field of characteristics $0$. The Hermitian structure
played no role yet.

Let us note first that the S-rank is associated with a point in a
projective space rather than with a tensor itself. Hence, we can
restrict considerations to tensors of length $1$ (as vectors in
the Hilbert space
$\mathcal{H}_1\otimes\cdots\otimes\mathcal{H}_k$). According to
Theorem~\ref{theo:simple1}, a tensor
$u\in\mathcal{H}_1\otimes\cdots\otimes\mathcal{H}_k$ is simple if
and only if $u\otimes u=\sigma_i(u\otimes u)$ for all $i=1,\ldots,
k$, where $\sigma_i$ interweaves the two copies of $\mathcal{H}_i$
in $\mathcal{H}_1\otimes\cdots\otimes\mathcal{H}_k
\otimes\mathcal{H}_1\otimes\cdots\otimes\mathcal{H}_k$. Note that
$\sigma_i$ acts as a unitary operator in
$(\mathcal{H}_1\otimes\cdots\otimes\mathcal{H}_k)^{\ot 2}$. This
means, in turn, that
\begin{equation}\label{}
\sum_{i=1}^k ||\,u\otimes u-\sigma_i(u\otimes u)||^2=0.
\end{equation}
The latter we can write as
\begin{equation}\label{}
\sum_{i=1}^k \langle u\otimes u-\sigma_i(u\otimes u)| u\otimes
u-\sigma_i(u\otimes u)\rangle=0,
\end{equation}
which, for tensors of length $1$ is equivalent to $ \sum_{i=1}^k
\mathrm{Re}\langle u\otimes u|\sigma_i(u\otimes u)\rangle=k$, or,
finally, to
$$ \mathrm{Re}\left\langle u\otimes u\,|\left(\sum_{i=1}^k
\sigma_i\right) (u\otimes u)\right\rangle=k\,.$$ The Schwarz
inequality yields now $u\otimes u=\bar\sigma(u\otimes u)$, where
$\bar\sigma=\frac{1}{k}\sum_{i=1}^n\sigma_i$, i.e.,
$$
\bar\sigma(x_1\otimes\cdots\otimes x_k\otimes
y_1\otimes\cdots\otimes y_k) =\frac{1}{k}\sum_{i=1}^n
x_1\otimes\cdots\otimes y_i\otimes\cdots x_k\otimes
y_1\otimes\cdots\otimes x_i\otimes \cdots\otimes y_k\,.
$$
In this way we have proven the following.
\begin{theo}\label{theo:simple2} Let
$u\in\mathcal{H}_1\otimes\cdots\otimes\mathcal{H}_k$, $||u||=1$.
Then, $u$ is simple if and only if
$$
\langle u\otimes u,\bar\sigma(u\otimes u)\rangle=1\,.
$$
\end{theo}
For symmetric tensors, a similar fact can be proven analogously.
\begin{theo}\label{theo:simple2s}
Let $u\in\mathcal{H}^{\vee k}$, $||u||=1$. Then, $u$ is simple
symmetric if and only if
$$
\langle u\otimes u,\sigma_k(u\otimes u)\rangle=1\,.
$$
\end{theo}

The structure of a Hilbert space $\mathcal{H}$ has been used in
\cite{schliemann01,sckll01,eckert02} to define simple symmetric
and simple antisymmetric tensors for tensors of order $2$ by means
of the \textit{Slater decomposition}. Both decompositions are
direct consequences of a possibility to write a complex symmetric
(antisymmetric) matrix in a diagonal (block-diagonal) form by a
unitary change of basis (Takagi theorem \cite{horn85}). The
existence of the Slater decompositions means that any symmetric
tensor $v\in\mathcal{H}\vee\mathcal{H}$ and any antisymmetric
tensor $w\in\mathcal{H}\wedge\mathcal{H}$ can be written as
\begin{eqnarray}\label{takagi}
v&=&\sum_{i=1}^n \lambda_i\, e_i\vee e_i, \quad \lambda_i> 0, \\
w&=&\sum_{i=1}^n \lambda_i\, f_i\wedge f_{n+i}, \quad \lambda_i>
0,
\end{eqnarray}
for some orthonormal systems $(e_i)$ and $(f_i)$ of $\mathcal{H}$.
The {\it Slater rank} is the number $n$ of terms in these
decompositions. The above are clearly symmetric and antisymmetric
analogues of the {\it Schmidt decomposition}: any tensor
$u\in\cH_1\otimes\cH_2$ can be written in the form
\begin{equation}\label{schm} u=\sum_{i=1}^n \lambda_i\, e_i\otimes
f_i, \quad \lambda_i> 0\,,
\end{equation}
for some orthonormal systems: $(e_i)$ in $\cH_1$, and $(f_i)$ in
$\mathcal{H}_2$.

\begin{theo}\label{srank} For any 2-tensor $u\in\cH_1\otimes\cH_2$
its Schmidt rank equals its S-rank.
\end{theo}
\textbf{Proof:} It is clear that the Schmidt decomposition
(\ref{schm}) implies that the S-rank of $u$ is $n$. Conversely, if
the S-rank of $u$ is $n$, then $u$ can be written in the form
(\ref{schm}) with the only difference that the systems $(e_i)$ and
$(f_i)$ are merely linearly independent. But this is a standard
procedure, used in the proof of existence of the Schmidt
decomposition, that we can choose $(e_i)$ and $(f_i)$ orthonormal.

$\Box$

 Hence, the 2-tensors are simple (resp., simple
symmetric, simple antisymmetric), if there exists a Schmidt (resp.
Slater) decomposition with a single $\lambda_i>0$, i.e. they have
the Schmidt (Slater) rank 1. Unfortunately, there are no direct
analogues of these decompositions for tensors of higher orders.

On the other hand, as we have already seen, such type of a
decomposition is not necessary to define (and check) which tensors
are simple symmetric (resp., simple antisymmetric), as the S-rank
can serve in these cases.

\section{Entanglement for multipartite Bose and Fermi systems}
\label{sec:quadratic}

Using the concept of simple tensors we can define simple
(non-entangled or separable) and entangled pure states for
multipartite systems of bosons and fermions.
\begin{definition}
\begin{description}\

\item{(a)} A pure state $\rho_v$ on $\mathcal{H}^{\vee k}$ (resp.,
on $\mathcal{H}^{\we k}$), $\rho_v=\frac{\kb{v}{v}}{||v||^2}$,
with $v\in\mathcal{H}^{\vee k}$ (resp., $v\in\mathcal{H}^{\we
k}$), $v\ne 0$, is called a \emph{bosonic} (resp.,
\emph{fermionic}) {\it simple} (or {\it non-entangled}) {\it pure
state}, if $v$ is a simple symmetric (resp., antisymmetric)
tensor. If $v$ is not simple symmetric (resp., antisymmetric), we
call $\rho_v$ a \emph{bosonic} (resp., \emph{fermionic})
\emph{entangled state}.

\item{(b)} A mixed state $\rho$ on $\mathcal{H}^{\vee k}$ (resp., on
$\mathcal{H}^{\we k}$) we call \emph{bosonic} (resp.,
\emph{fermionic}) {\it simple} (or {\it non-entangled}) {\it mixed
state}, if it can be written as a convex combination of {bosonic}
(resp., {fermionic}) simple pure states. In the other case $\rho$
is called \emph{bosonic} (resp., \emph{fermionic}) \emph{entangled
mixed state}.
\end{description}
\end{definition}

According to Theorem \ref{theo:simple}, bosonic simple pure
$k$-states are of the form
$$\kb{e{\vee}\cdots\vee e}{e{\vee}\cdots\vee e}$$
for some unit vector $e\in\cH$, and fermionic simple pure
$k$-states are of the form
$${k!}\kb{e_1\we\cdots\we e_k}{e_1\we\cdots\we e_k}$$ for some
orthonormal system $e_1,\dots,e_k$ in $\cH$.

Fixing a base in $\mathcal{H}$ results in defining coefficients
$[u^{i_1\ldots i_k}]$ of $u\in\mathcal{H}^{\otimes k}$. Formulae
characterizing simple tensors, thus simple pure states, can be
written in forms of quadratic equations with respect to these
coefficients as follows. The corresponding characterization of
entangled pure states are obtained by negation of the latter.

\begin{theo}\label{theo:quadratic1} The pure state $\rho_u$,
associated with a tensor $u=[u^{i_1\ldots
i_k}]\in\mathcal{H}^{\otimes k}$, is entangled if and only if
there exist $i_1,\ldots,i_k,j_1,\ldots,j_k$, and  $s=1,\ldots,k$
such that
\begin{equation}\label{quadratic}
u^{i_1\ldots i_s\ldots i_k}u^{j_1\ldots j_s\ldots j_k}\ne
u^{i_1\ldots j_s\ldots i_k}u^{j_1\ldots i_s\ldots j_k}.
\end{equation}

\end{theo}
\textbf{Proof:} The tensor $u\otimes u$ has coefficients
$u^{i_1\ldots i_k} u^{j_1\ldots k_k}$, so Eq.(\ref{quadratic})
expresses the fact that $u\otimes u\ne\sigma_s(u\otimes u)$, and
thus Theorem~\ref{theo:quadratic1} is a direct consequence of
Theorem~\ref{theo:simple1}.

$\Box$

\begin{theo}\label{theo:quadratic2}
The bosonic pure state $\rho_v$, associated with a symmetric
tensor $v=[v^{i_1\ldots i_k}]\in\mathcal{H}^{\vee k}$, is bosonic
entangled if and only if there exist $i_1,\ldots,i_k,j_1,\ldots,j_k$, such that
\begin{equation}\label{quadratic2}
v^{i_1\ldots i_{k-1}i_k}v^{j_1\ldots j_{k-1}j_k}\ne v^{i_1\ldots
i_{k-1}j_k}v^{j_1\ldots j_{k-1}i_k}\,.
\end{equation}
\end{theo}
\textbf{Proof:} A direct consequence of
Theorem~\ref{theo:simplesym}.

$\Box$

\begin{example}
The bosonic pure state associated with the symmetric tensor $u$ defined in (\ref{u0}) is entangled, as we know it has the S-rank 2. To apply our characterization (\ref{quadratic2}), note that the only non-zero coefficients are $u^{12}=u^{21}=\frac{1}{2}$. Hence,
$$0=u^{11}u^{22}\ne u^{12}u^{21}=\frac{1}{4}\,.$$
\end{example}

\begin{example}
The bosonic pure state associated with the symmetric tensor $u$ defined in (\ref{u}) is non-entangled, as we know it has the S-rank 1. To apply our characterization (\ref{quadratic2}), note that, as easily checked, $u^{i_1i_2i_3i_4}=2^{i(3)}$, where $i(3)$ is the number of 3's in the sequence  $i=(i_1,i_2,i_3,i_4)$. Now it is clear that
$$u^{i_1i_2i_3i_4}u^{j_1j_2j_3j_4}=2^{i(3)+j(3)}\,,$$
where $i(3)+j(3)$ is the number of 3's in the sequence $(i_1,i_2,i_3,i_4,j_1,j_2,j_3,j_4)$ which remains unchanged under permutations. In particular,
$$u^{i_1i_2i_3i_4}u^{j_1j_2j_3j_4}=u^{i_1i_2i_3j_4}u^{j_1j_2j_3i_4}\,.$$
\end{example}

\begin{theo}\label{theo:quadratic3}
The fermionic pure state $\rho_w$, associated with an
antisymmetric tensor $w=[w^{i_1\ldots i_k}]\in\mathcal{H}^{\wedge
k}$, is fermionic entangled if and only if there exist
$i_1,\ldots,i_{k+1},j_1,\ldots,j_{k-1}$ such that
\begin{equation}\label{quadratic3}
w^{[i_1\ldots i_{k}}w^{i_{k+1}]j_1\ldots j_{k-1}}\ne 0\,,
\end{equation}
where the left-hand side is the antisymmetrization of $w^{i_1\ldots i_{k}}w^{i_{k+1}j_1\ldots j_{k-1}}$ with respect to indices $i_1,\dots,i_{k+1}$.
\end{theo}
\textbf{Proof:} A direct consequence of
Theorem~\ref{theo:simpleantisym}.

$\Box$

In view of the above characterizations, it is obvious that the
sets of entangled (entangled bosonic, entangled fermionic) pure
states are open: pure states sufficiently close to entangled ones
are entangled.

\begin{example}
The fermionic pure state associated with the antisymmetric tensor $w$ defined in (\ref{w}) is non-entangled, as we know its S-rank is 2. To apply our characterization (\ref{quadratic3}), let us note that the only non-zero coefficients are $w^{12}=-w^{21}=w^{23}=-w^{32}=\frac{1}{2}$. The antisymmetrization $w^{[i_1i_2}w^{i_3]j}$ can be non-zero only if $(i_1,i_2,i_3)$ is a permutation of $(1,2,3)$. But, as easily seen,
$$w^{[12}w^{3]j}=\frac{1}{3}\left(w^{12}w^{3j}-w^{32}w^{1j}\right)=0$$
for $j=1,2,3$.
\end{example}

\begin{example}
The fermionic pure state associated with the antisymmetric tensor $w$ defined in (\ref{w1}) is entangled, as we know its S-rank is 4. To apply our characterization (\ref{quadratic3}), let us note that the only non-zero coefficients are $w^{12}=-w^{21}=w^{34}=-w^{43}=\frac{1}{2}$. As easily seen,
$$w^{[12}w^{3]4}=\frac{1}{3}w^{12}w^{34}=\frac{1}{12}\ne 0\,.$$
\end{example}

\section{Jamio{\l}kowski isomorphisms for bi-partite systems of bosons and fermions}
\label{sec:jamiol}

In the theory of entanglement there exists a useful tool for investigating,
on one hand, entanglement properties of states and, on the other, structure
of positivity-preserving and hermiticity-preserving maps between matrix
algebras called the Jamio{\l}kowski isomorphism \cite{bengtsson06}.
Originally proposed in \cite{jamiolkowski72} as an instrument for checking
the property of preserving positive semi-definiteness for a linear map
between two matrix algebras in finite dimensional spaces, it was later used to
prove the so-called operator form representation for linear maps on quantum
states (i.e.\ positive semi-definite operators) \cite{choi75}. The
construction can be extended to the infinite-dimensional setting if the
restriction to Hilbert-Schmidt operators is imposed \cite{grabowski07}) which
significantly heightens the usefulness of the Jamio{\l}kowski isomorphism in
cases when finite-dimensional description of q quantum system is not
suitable. An in-depth description of the applicability the Jamio{\l}kowski
isomorphism is given in \cite{arighi04} where various properties of
entanglement and separability of states are paralleled with features of the
corresponding linear maps.

One of the most straightforward applications of the Jamio{\l}kowski
isomorphism is a characterization of separability of a state in terms of the
rank of the corresponding map. In the following we show how this decription
can be extended to states of indistinguishable particles.

Let us recall (see e.g. \cite{grabowski07}) that for two Hilbert spaces
$\cH_1$ and $\cH_2$ we have the following diagram consisting of {\it
Jamio{\l}kowski isomorphisms}:
\begin{equation}\label{dam1} \xymatrix{
\cL_2(\cL_2(\cH_2),\cL_2(\cH_1)) \ar[ddrr]_{\cJ_1}
 \ar[rrr]^{{\cJ}}
 & &  & \cL_2(\cL_2(\cH_2,\cH_1))\ar[ddl]^{\cJ_2}
 \\ \\
   & & \cL_2(\cH_1\ot\cH_2) & & }
\end{equation}
Here, with $\cL_2(\cH_1,\cH_2)$ we denote the Hilbert space of
Hilbert-Schmidt maps from $\cH_1$ into $\cH_2$. Of course, this
space reduces to all complex linear maps, if $\cH_1$ or $\cH_2$ is
finite-dimensional. Note that we write shorter $\cL_2(\cH)$ for
$\cL_2(\cH,\cH)$. Note also that according to the obvious
identification $\cL_2(\cH_2,\cH_1)\simeq \cH_1\ot\cH_2^*$ and the
identification
\begin{equation}\label{id2} \mathcal{L}\left(
\mathcal{H}_2,\mathcal{H}_1\right)^\ast \simeq\mathcal{L}\left(
\mathcal{H}_1,\mathcal{H}_2\right)\,,
\end{equation}
induced by the natural pairing
\begin{equation}\label{id3}
\mathcal{L}\left(\mathcal{H}_2,\mathcal{H}_1\right)
\ti\mathcal{L}\left( \mathcal{H}_1,\mathcal{H}_2\right)\,\ni
(A,B)\mapsto \tr(A\circ B)\in\C\,,
\end{equation}
we can rewrite the above diagram in the form
\begin{equation}\label{dam}
\xymatrix{ \cH_1\ot\cH_1^*\ot\cH_2\ot\cH_2^* \ar[ddr]_{\cJ_1}
 \ar[rr]^{{\cJ}}
    && \cH_1\ot\cH_2^*\ot\cH_2\ot\cH_1^*\ar[ddl]^{\cJ_2}
 \\ \\
    & \cH_1\ot\cH_2\ot\cH_2^*\ot\cH_1^* &}
\end{equation}
in which the Jamio\l kowski isomorphisms reduce to appropriate
permutations of tensors. In the case when $\cH_1=\cH_2$ the whole
picture reduces to
\begin{equation}\label{dam2} \xymatrix{
\cL_2(\cL_2(\cH)) \ar[ddrr]_{\cJ_1}
 \ar[rrr]^{{\cJ}}
 & &  & \cL_2(\cL_2(\cH))\ar[ddl]^{\cJ_2}
 \\ \\
   & & \cL_2(\cH\ot\cH) & & }\,.
\end{equation}
We can now decompose $\cH\ot\cH$ into
$(\cH\vee\cH)\oplus(\cH\we\cH)$ and reduce our diagram to maps on
each of these components. In this way we get
\begin{equation}\label{dam3} \xymatrix{
\cL_2^{sa}(\cH\otimes\cH^*) \ar[ddrr]_{\cJ_1}
 \ar[rrr]^{{\cJ}}
 & &  & \cL_2^{sa}(\cH\otimes\cH^*)\ar[ddl]^{\cJ_2}
 \\ \\
   & & \cL_2(\cH^{\vee 2})\oplus\cL_2(\cH^{\we 2}) & & }\,.
\end{equation}
Here, $\cL_2^{sa}(\cH\otimes\cH^*)$ is the space of self-adjoint
Hilbert-Schmidt operators $\Phi$, $\Phi=\Phi^*$, on
$\cH\otimes\cH^*=\cL_2(\cH)$ according to the identification
$\cL_2(\cH)\simeq(\cL_2(\cH))^*$ (see (\ref{id2})) related with
the pairing (\ref{id3}). Note that this is different from
hermicity, $\Phi^*\ne\Phi^\dag$, since $\Phi^\dag$ depends on
$\Phi$ anti-linearly.

Indeed, $\Phi$ written in the Dirac notation as
$\Phi=\lambda_{ijkl}\ket{e_i}\ot\bra{e_j}\ot\ket{e_k}\ot\bra{e_l}$
is self-adjoint if and only if
$\tr(A\circ\Phi(B))=\tr(\Phi(A)\circ B)$ for all
$A,B\in\cL_2(\cH)$, that applied to $A=\ket{e_j}\ot\bra{e_i}$ and
$B=\ket{e_l}\ot\bra{e_k}$ yields $\lambda_{ijkl}=\lambda_{klij}$,
so that we deal with maps coming from symmetric or antisymmetric
tensors. What is more, since
$${\cJ}\left(\ket{e_i}\ot\bra{e_j}\ot\ket{e_k}\ot\bra{e_l}\right)=\ket{e_i}\ot\bra{e_l}\ot\ket{e_k}\ot\bra{e_j}\,,$$
we have a further splitting
$$\cL_2^{sa}(\cH\otimes\cH^*)=\cL_2^{sa+}(\cH\otimes\cH^*)\oplus\cL_2^{sa-}(\cH\otimes\cH^*)\,,$$
where $\cL_2^{sa\pm}(\cH\otimes\cH^*)$ consists of these $\Phi$
for which ${\cJ}(\Phi)=\pm\Phi$. Finally, we end up with {\it
bosonic} and {\it fermionic Jamio\l kowski maps}:
\begin{equation}\label{dam4}
\xymatrix{ \cL_2^{sa+}(\cH\otimes\cH^*) \ar[ddrr]_{\cJ_1^+}
 \ar[rrr]^{{\cJ}^+}
 & &  & \cL_2^{sa+}(\cH\otimes\cH^*)\ar[ddl]^{\cJ_2^+}
 \\ \\
   & & \cL_2(\cH^{\vee 2}) & & }
\end{equation}
and
\begin{equation}\label{dam5}
\xymatrix{ \cL_2^{sa-}(\cH\otimes\cH^*) \ar[ddrr]_{\cJ_1^-}
 \ar[rrr]^{{\cJ}^-}
 & &  & \cL_2^{sa-}(\cH\otimes\cH^*)\ar[ddl]^{\cJ_2^-}
 \\ \\
   & & \cL_2(\cH^{\we 2}) & & }\,.
\end{equation}

If now $\rho=\kb{v}{v}$ is a pure state in $\cH^{\vee 2}$
corresponding to a vector $v\in\cH^{\vee 2}$ with a Slater
decomposition $v=\sum_{i=1}^r\lambda_ie_i\vee e_i$, $\lambda_i>0$,
so that the Slater rank is $r$, then $\rho=\cJ_1^+(\Phi)$, with
$$\Phi=\sum_{i,j}\lambda_i\lambda_j\ket{e_i}\ot\bra{e_j}\ot\ket{e_i}\ot\bra{e_j}$$
being a map from $\cL_2^{sa+}(\cH\otimes\cH^*)$ of rank $r^2$.

Similarly, if $\rho=\kb{w}{w}$ is a pure state in $\cH^{\we 2}$
corresponding to a vector $w\in\cH^{\vee 2}$ with a Slater
decomposition $w=\sum_{i=1}^r\mu_if_i\we f_{n+i}$, $\mu_i>0$, so
that the Slater rank is $r$, then $\rho=\cJ_1^- (\Phi)$, with
\begin{eqnarray*}\Phi&=&\sum_{i,j}\mu_i\mu_j\left(\ket{f_i}\ot\bra{f_j}\ot\ket{f_{n+i}}\ot\bra{f_{n+j}}-
\ket{f_{n+i}}\ot\bra{f_j}\ot\ket{f_{i}}\ot\bra{f_{n+j}}\right.\\
&&-\left.\ket{f_i}\ot\bra{f_{n+j}}\ot\ket{f_{n+i}}\ot\bra{f_{j}}-
\ket{f_{n+i}}\ot\bra{f_{n+j}}\ot\ket{f_{i}}\ot\bra{f_{j}}\right)
\end{eqnarray*} being a map from $\cL_2^{sa-}(\cH\otimes\cH^*)$ of
rank $4r^2$. In this way we get a characterization of bosonic and
fermionic simple pure states in terms of the corresponding Jamio\l
kowski isomorphisms.

\begin{theo} A pure state $\rho$ in $\cH^{\vee 2}$ (resp., $\cH^{\we
2}$) is bosonic (resp., fermionic) simple if and only if
$\rho=\cJ_1^+(\Phi)$ (resp., $\rho=\cJ_1^-(\Phi)$) for
$\Phi\in\cL_2^{sa+}(\cH\otimes\cH^*)$ (resp.,
$\Phi\in\cL_2^{sa-}(\cH\otimes\cH^*)$) of rank 1 (resp., 4).
\end{theo}
\begin{remark} Of course, choosing a basis in $\cH$ we can
represent the above maps by matrices and gets the Jamio\l kowski
isomorphism in the form of permutation of indices, more familiar
to physicists. The above form has the advantage that it does not
depend on the basis, i.e. is canonical and covariant.
\end{remark}

\section{Entangled states of composite systems
with generalized parastatistics} \label{sec:segrepara}

Our approach to the entanglement of composite systems for
identical particles is so general and natural that it allows for
an immediate implications also for generalized parastatistics.

Observe first that simple tensors of length $1$ in
$\tilde{\mathcal{H}}=\mathcal{H}_1\otimes\cdots\otimes\mathcal{H}_k$
form an orbit of the group $U(\mathcal{H}_1)\times\cdots\times
U(\mathcal{H}_k)$ acting on $\tilde{\mathcal{H}}$ in the obvious
way. In fact, each such tensor can be written as
$e_1^1\otimes\cdots\otimes e_1^k$ for certain choice of
orthonormal bases $e_1^j,\dots,e_{n_j}^j$ in $\mathcal{H}_j$,
$j=1,\ldots,k$. This means that simple tensors are just vectors of
highest (or lowest -- depending on the convention) weight of the
compact Lie group $U(\mathcal{H}_1)\times\cdots\times
U(\mathcal{H}_{k})$ relative to some choice of a maximal
torus and Borel subgroups. If indistinguishable particles are
concerned, the symmetric and antisymmetric tensors in
$\mathcal{H}^k$ form particular irreducible parts of the
`diagonal' representation of the compact group $U(\mathcal{H})$ in
the Hilbert space $\mathcal{H}^{\otimes k}$, defined by
\begin{equation}\label{trep}
U(x_1\otimes\cdots\otimes x_k)=U(x_1)\otimes\cdots\otimes U(x_k).
\end{equation}
Recall that we identify the symmetry group $S_k$ with the group of
certain unitary operators on the Hilbert space $\mathcal{H}^k$ in
the obvious way,
$$\sigma(x_1\otimes\cdots\otimes x_k)=x_{\sigma(1)}\otimes\cdots\otimes x_{\sigma(k)}\,.$$
Note that the operators of $S_k$ intertwine the unitary action of
$U(\mathcal{H})$. In the cases of the symmetric and antisymmetric
tensors, we speak about Bose and Fermi statistics, respectively.
But, for $k>2$, there are other irreducible parts of the
representation (\ref{trep}), associated with invariant subspaces
of the $S_k$-action, that we shall call {\it (generalized)
parastatistics}. Any of these $k$-parastatistics (i.e. any
irreducible subspace of the tensor product $\mathcal{H}^{\otimes
k}$) is associated with a {\it Young tableau} $\alpha$ with
$k$-boxes (chambers) as follows (see e.g. \cite{fulton91,fulton97}).

Consider partitions of $k$: $k=\lambda_1+\cdots+\lambda_r$, where
$\lambda_1\ge\cdots\ge\lambda_r\ge 1$. To a partition
$\lambda=(\lambda_1,\dots,\lambda_r)$ is associated a {\it Young
diagram} (sometimes called a {\it Young frame} or a {\it Ferrers
diagram}) with $\lambda_i$ boxes in the $i$th row, the rows of
boxes lined up on the left. Define a {\it tableau} on a given
Young diagram to be a numbering of the boxes by the integers
$1,\dots,k$, and denote with $Y_\lambda$ the set of all such Young
tableaux. Finally, put $Y(k)$ to be the set of all Young tableaux
with $k$ boxes. Given a tableau $\alpha\in Y(k)$ define two
subgroups in the symmetry group $S_k$:
$$P=P_\alpha=\{\sigma\in S_k: \sigma \text{\ preserves each row of}\ \alpha\}$$
and
$$Q=Q_\alpha=\{\sigma\in S_k: \sigma \text{\ preserves each column of}\ \alpha\}\,.$$
In the space of linear operators on $\mathcal{H}^{\ot k}$ we
introduce two operators associated with these subgroups:
\begin{equation}\label{yoper}
a_\alpha=\sum_{\sigma \in P}\sigma\,,\quad b_\alpha=\sum_{\sigma
\in Q}(-1)^{\sigma}\sigma\,.
\end{equation}
Finally, we define the {\it Young symmetrizer}
\begin{equation}\label{ysym}
c_\alpha=a_\alpha\circ b_\alpha=\sum_{\tau \in P,\, \sigma\in
Q}(-1)^{\sigma}\tau\circ\sigma\,.
\end{equation}
It is well known that $\pi^\alpha=\frac{1}{\mu(\alpha)}c_\alpha$,
for some non-zero rational number $\mu(\alpha)$, is an orthogonal
projector and that the image $\mathcal{H}^\alpha$ of $c_\alpha$ is
an irreducible subrepresentation of $U(\mathcal{H})$, i.e. the
parastatistics associated with $\alpha$. As a matter of fact,
these representations for Young tableaux on the same Young diagram
are equivalent, so that the constant $\mu(\alpha)$ depends only on
the Young diagram $\lambda$ of $\alpha$ (does not depend on the
enumeration of boxes), $\mu(\alpha)=\mu(\lambda)$, and is related
to the multiplicity $m(\lambda)$ of this irreducible
representation in $\mathcal{H}^{\ot k}$ by $\mu(\lambda)\cdot
m(\lambda)=k!$. For a given Young diagram (partition) $\lambda$,
the map
\begin{equation}\label{cysym}
\epsilon_\lambda=\frac{1}{\mu(\lambda)^2}\sum_{\alpha\in
Y_\lambda} c_\alpha
\end{equation}
is an orthogonal projection, called the {\it central Young
symmetrizer}, onto the invariant subspace being the sum of all
copies of the irreducible representations equivalent to that with
a parastatistics from $Y_\lambda$.

The symmetrization $\pi^\vee$ (antisymmetrization $\pi^\wedge$)
projection corresponds to a Young tableau with just one row (one
column) and arbitrary enumeration. It is well known that any
irreducible representation $\mathcal{H}^\alpha$ of
$U(\mathcal{H})$ contains cyclic vectors which are of highest
weight relative to some choice of a maximal torus and Borel
subgroups in $U(\mathcal{H})$. We will call them
\textit{$\alpha$-simple vectors} or \textit{simple vectors in
$\mathcal{H}^\alpha$}. Note that such vectors can be viewed as
\textit{generalized coherent states} \cite{perelomov86}. They were
also regarded as the `most classical' states by several authors
\cite{kb09}. These are exactly the tensors associated with simple
(non-entangled) pure states for composite systems of particles
with (generalized) parastatistics. This is because $\alpha$-simple
tensors represent the minimal amount of quantum correlations for
tensors in $\mathcal{H}^\alpha$: the quantum correlations forced
directly by the particular parastatistics.
\begin{example}\label{ex}
\begin{description}
\item{(a)} For $k=2$ we have just the obvious splitting of
    $\mathcal{H}^{\otimes 2}$ into symmetric and antisymmetric tensors:
    $\mathcal{H}^{\wedge 2}\oplus\mathcal{H}^{\vee 2}$.
\item{(b)} For $k=3$, besides symmetric and antisymmetric tensors,
we have two additional irreducible parts associated
    with the Young tableaux
\end{description}
\begin{equation}\label{}
\Yvcentermath1 \alpha_1=\young(12,3) \quad and \quad
\alpha_2=\young(13,2)\ ,
\end{equation}
respectively. Hence,
\begin{equation}\label{}
\mathcal{H}^{\otimes 3}=\mathcal{H}^{\wedge 3}\oplus
\mathcal{H}^{\alpha_1}\oplus\mathcal{H}^{\alpha_2}\oplus
\mathcal{H}^{\vee 3}\,,
\end{equation}
with
\begin{equation}\label{}
\pi^{\alpha_1}:\mathcal{H}^{\otimes
3}\rightarrow\mathcal{H}^{\alpha_1},
\end{equation}
\begin{equation}
\pi^{\alpha_1}(x_1\otimes x_2\otimes x_3)=\frac{1}{3} (x_1\otimes
x_2\otimes x_3 +x_2\otimes x_1\otimes x_3 -x_3\otimes x_2\otimes
x_1 -x_3\otimes x_1\otimes x_2),
\end{equation}
and
\begin{equation}\label{}
\pi^{\alpha_2}:\mathcal{H}^{\otimes
3}\rightarrow\mathcal{H}^{\alpha_2},
\end{equation}
\begin{equation}
\pi^{\alpha_2}(x_1\otimes x_2\otimes x_3)=\frac{1}{3} (x_1\otimes
x_2\otimes x_3 +x_3\otimes x_2\otimes x_1 -x_2\otimes x_1\otimes
x_3 -x_2\otimes x_3\otimes x_1).
\end{equation}
\end{example}
The simple tensors (the highest weight vectors) in
$\mathcal{H}^{\alpha_1}$ can be written as
\begin{equation}\label{}
v^{\alpha_1}_\lambda=\lambda(e_1\otimes e_1\otimes e_2-e_2\otimes
e_1 \otimes e_1),
\end{equation}
for certain choice of an orthonormal basis $e_i$ in $\mathcal{H}$
and $\lambda\ne 0$. Analogously, the simple tensors in
$\mathcal{H}^{\alpha_2}$, in turn, take the form
\begin{equation}\label{}
v^{\alpha_2}_\lambda=\lambda(e_1\otimes e_2\otimes e_1-e_2\otimes
e_1 \otimes e_1).
\end{equation}
For $\text{dim}(\mathcal{H})=3$, the simple tensors of length $1$
form an orbit of the unitary group $U(\mathcal{H})$ of the (real)
dimension $7$ in $\mathcal{H}^{\alpha_1}$ and
$\mathcal{H}^{\alpha_2}$. The simple symmetric tensors of length
$1$ form an orbit of the dimension $5$, and the simple
antisymmetric ones (of length $1$) -- an orbit of the dimension 1.
The dimensions of the irreducible representations are:
$\text{dim}(\mathcal{H}^{\we 3})=1$,\
$\text{dim}(\mathcal{H}^{\vee 3})=10$,\
$\text{dim}(\mathcal{H}^{\alpha_1})=\text{dim}(\mathcal{H}^{\alpha_2})=8$\,.

\bigskip
Let $\mathcal{H}^\alpha\subset\mathcal{H}^{\otimes k}$ denotes the
irreducible component of the tensor representation of the unitary
group $U(\mathcal{H})$ in $\mathcal{H}^{\otimes k}$ associated
with a Young diagram $\alpha\in Y(k)$.
\begin{definition}\

\begin{description}
\item{(a)} We say that a pure state $\rho\in\mathcal{H}^{\otimes k}$ {\it
obeys a parastatistics} $\alpha\in Y(k)$ (is a {\it pure
$\alpha$-state} in short) if $\rho$ {\it is represented} by a
nonzero tensor $v\in\mathcal{H}^\alpha$, i.e.
\begin{equation}\label{rho}\rho=\rho_v=\frac{\kb{v}{v}}{||v||^2}\,.
\end{equation}
In other words, $\rho$ is a pure state on the Hilbert space
$\cH^\alpha$.
\item{(b)} A pure state $\rho\in\mathcal{H}^{\otimes k}$,
obeying a parastatistics $\alpha$ is called a \emph{simple pure
state for the parastatistics} $\alpha$ (\emph{simple pure
$\alpha$-state}, in short), if $\rho$ is represented by an
$\alpha$-simple tensor in $\mathcal{H}^\alpha$. If $\rho$ is not
simple $\alpha$-state, we call it \emph{entangled pure
$\alpha$-state}.
\item{(c)} A mixed state $\rho$ on $\mathcal{H}^{\alpha}$ we call
a \emph{simple (mixed) state for the parastatistics} $\alpha$
(\emph{simple $\alpha$-state} in short), if it can be written as a
convex combination of  simple pure $\alpha$-states. In the other
case, $\rho$ is called \emph{entangled mixed $\alpha$-state}.
\end{description}
\end{definition}

For an arbitrary parastatistics (Young tableau)
$\alpha\in Y(k)$ with the partition (Young diagram)
$\lambda=(\lambda_1,\dots,\lambda_r)$, consider the map
$$i_\alpha:\mathcal{H}^{\times r}\rightarrow \mathcal{H}^{\otimes k}\,,
\quad (x_1,\dots,x_r)\mapsto x_{\alpha(1)}\otimes\cdots\otimes
x_{\alpha(k)}\,,$$ where $\alpha(i)$ is the number of the raw in
which the box with the number $i$ appears in the tableaux
$\alpha$. In other words, we make a tensor product of $k$ vectors
from $\{ x_1,\dots,x_r\}$ by putting $x_j$ in the places indicated
by the number of the boxes in the $j$th row. For instance, the
Young tableaux from Example \ref{ex} give
$i_{\alpha_1}(x_1,x_2)=x_1\otimes x_1\otimes x_2$ and
$i_{\alpha_2}(x_1,x_2)=x_1\otimes x_2\otimes x_1$. It is clear
that $i_\alpha(x_1,\dots,x_r)$ is an eigenvector of $a_\alpha$.

It is easy to see now that the S-rank of the tensor
$\pi^\alpha(x_{\alpha(1)}\otimes\cdots\otimes x_{\alpha(k)})$, for
$x_1\we\cdots\we x_r\ne 0$, is $r$ and that this is the minimal
S-rank for tensors from $\cH^\alpha$. Hence, the minimality of the
S-rank is a good characteristic also for simple $\alpha$-tensors.
\begin{theo} A tensor $u\in\cH^\alpha$ is $\alpha$-simple if and
only if $u$ has the minimal S-rank among all non-zero tensors from
$\cH^\alpha$. This minimal S-rank equals $r$ -- the number of rows
in the corresponding Young diagram.
\end{theo}

\section{Conclusions}

States of identical particles exhibit \textit{a priori}
correlations caused merely by (anti)symmetry of the wave function
in the case of fermions or bosons. It is thus reasonable to treat
as an analogue of the entanglement encountered in systems of
distinguishable particles only an additional amount of correlation
going beyond that stemming from symmetry requirements.

We proposed a way of treating all non-classical correlation, i.e.,
those which can be identified with the `genuine entanglement' and
not caused merely by symmetries. This unifies all cases: of
distinguishable particles, fermions, and bosons, and can be easily
extended to hypothetical multipartite systems consisting of
particles subjected to arbitrary parastatistics.

We defined simple (non-entangled) pure states as one-dimensional
selfadjoint projectors associated with simple tensors obeying
appropriate symmetries identifying particles as bosons, fermions,
etc. Consequently, simple (non-entangled) mixed states were
defined as convex combinations of  simple (non-entangled) pure
ones. Such an unifying approach allowed also for description of
such tools, known from the entanglement
theory, as the Jamio{\l}kowski isomorphism and Schmidt rank, to
systems with other symmetries. The introduced concept of {\it
S-rank} not only provides us with a tool for distinguishing
entanglement of pure states with a given parastatistics, but is
interesting also {\it per se}, as it offers the simplest
characterization of highest weight vectors we know.

In the case of two fermionic subsystems our approach identifies
non-entangled pure states to be the same as in all other
approaches mentioned in Introduction, i.e., we identify them with
simple antisymmetric tensors in the meaning explained in
Section~\ref{sec:c-rank} above. In the bosonic case, from the
geometric point of view, we clearly have {\it a priori} two
inequivalent types of non-entanglement: tensor products of
identical states and states associated with symmetrizations of products of orthogonal
vectors. Non-entangled states of two different types are not
connected by local unitary transformations which is in contrast to
the familiar situation of distinguishable particles and intuitions
build upon the fact that all separable states of distinguishable
particles can be obtained from a single one by local
transformations. Although this is obviously acceptable, it poses
an open fundamental problem what is a physical meaning of two
geometrically inequivalent types of non-entanglement.

In our approach we adopted the view that non-entangled pure
bosonic states are simple symmetric tensors - tensor products of
identical vectors. We find at least two arguments justifying this
choice. In \cite{paskauskas01} it was pointed that all states
which are symmetrizations of products of distinct vectors can be
used to perform such clearly `non-classical' tasks like
teleportation. This definitely remains in conflict with the basic
intuition connecting non-entanglement with the purely classical
world. Second, from purely mathematical point of view, only tensor
product of identical vectors provide highest weight vectors of the
corresponding representation of the unitary group, like in the
other cases.

Another achievements of our paper are explicit characterizations
of simplicity (non-entanglement) for pure states in terms of
directly verifiable, quadratic in coefficients, conditions which
are computationally much easier than those proposed in
the literature.


\begin{thebibliography}{10}

\bibitem{schrodinger35} E.~Schr\"odinger.
\newblock Discussion of probability relations between separated systems.
\newblock {\em Math. Proc. Cambridge Phil. Soc.}, 31:555--563, 1935.

\bibitem{herbut87} F~Herbut and M~Vuji\v{c}i\'c.
\newblock Irrelevance of the {P}auli principle in distant correlations between
  identical fermions.
\newblock {\em J. Phys. A}, 20(16):5555--5563, 1987.

\bibitem{grobe94} R~Grobe, K~Rz\c{a}\.zewski, and J~H Eberly.
\newblock {M}easure of electron-electron correlation in atomic physics.
\newblock {\em J. Phys. B}, 27(16):L503--L508, 1994.

\bibitem{schliemann01} J.~Schliemann, D.~Loss, and A.~H. MacDonald.
\newblock {D}ouble-occupancy errors, adiabaticity, and entanglement of spin
  qubits in quantum dots.
\newblock {\em Phys. Rev. B}, 63(8):085311, 2001.

\bibitem{sckll01} J.~Schliemann, J.~I. Cirac, M.~Ku\'s, M.~Lewenstein, and
    D.~Loss.
\newblock {Q}uantum correlations in two-fermion systems.
\newblock {\em Phys. Rev. A}, 64:022303, 2001.

\bibitem{eckert02} K.~Eckert, J.~Schliemann, D.~Bru{\ss}, and M~Lewenstein.
\newblock {Q}uantum correlations in systems of identical particles.
\newblock {\em Ann. Phys.}, 299:88--127, 2002.

\bibitem{paskauskas01} R.~Pa\v{s}kauskas and L.~You.
\newblock {Q}uantum correlations in two-boson wave functions.
\newblock {\em Phys. Rev. A}, 64(4):042310, 2001.

\bibitem{herbut01} F.~Herbut.
\newblock How to distinguish identical particles.
\newblock {\em Am. J. Phys.}, 69(2):207--217, 2001.

\bibitem{li01} Y.~S. Li, B.~Zeng, X.~S. Liu, and G.~L. Long.
\newblock {E}ntanglement in a two-identical-particle system.
\newblock {\em Phys. Rev. A}, 64(5):054302, 2001.

\bibitem{ghirardi02} G.~Ghirardi, L.~Marinatto, and T.~Weber.
\newblock {E}ntanglement and {P}roperties of {C}omposite {Q}uantum {S}ystems:
  {A} {C}onceptual and {M}athematical {A}nalysis.
\newblock {\em Journal of Statistical Physics}, 108(1):49--122, 2002.

\bibitem{ghirardi04} G.~Ghirardi and L.~Marinatto.
\newblock {G}eneral criterion for the entanglement of two indistinguishable
  particles.
\newblock {\em Phys. Rev. A}, 70(1):012109, 2004.

\bibitem{ghirardi05} G.~Ghirardi and L.~Marinatto.
\newblock {I}dentical particles and entanglement.
\newblock {\em Optics and Spectroscopy}, 99(3):386--390, 2005.

\bibitem{sasaki10} T.~Sasaki, T~Ichikawa and I.~Tsutsui.
\newblock {E}ntanglement of indistinguishable particles.
\newblock {\em Phys. Rev. A} {83}:012113, 2011.

\bibitem{grabowski05}
J. Grabowski, M. Ku\'s, and G. Marmo.
\newblock {G}eometry of quantum systems: density states and
entanglement.
\newblock  {\em J. Phys. A: Math. Gen.} {38}:10217--10244, 2005.

\bibitem{grabowski06}
J. Grabowski, M. Ku\'s, and G. Marmo.
\newblock {S}ymmetries, group actions, and entanglement.
\newblock {\em Open Sys. Information Dyn.}, {13}:343--362, 2006.

\bibitem{horn85} R.A. Horn and C.R. Johnson.
\newblock {Matrix analysis}.
\newblock Cambridge University Press, 1985.

\bibitem{bengtsson06} I. Bengtsson and K. \.Zyczkowski.
\newblock {G}eometry of {Q}uantum {S}tates.
\newblock Cambridge University Press, 2006.

\bibitem{jamiolkowski72} A. Jamio{\l}kowski.
\newblock {L}inear transformations which preserve trace and positive semidefinite
operators.
\newblock {\em Rep. Math. Phys.} {3}:275--278, 1972.

\bibitem{choi75} M.~D. Choi,
\newblock Completely positive linear maps on complex matrices.
\newblock {\em Linear Alg. Appl.} {10}:285-278, 1975.

\bibitem{grabowski07} J. Grabowski, M. Ku\'s, and G. Marmo.
\newblock {O}n the relation between states and maps in infinite
dimensions.
\newblock  {\em Open Sys. Information Dyn.},
{14}:355--370, 2006.

\bibitem{arighi04} P. Arrighi and C. Patricot.
\newblock {O}n quantum operations as quantum states.
\newblock {\em Ann. Phys.} {311}:26–-52, 2004.

\bibitem{fulton91} W. Fulton and J. Harris.
\newblock {R}epresentation Theory. A First Course.
\newblock Springer Verlag, 1991.

\bibitem{fulton97}
W. Fulton
\newblock {Y}oung Tableaux.
\newblock Cambridge University Press, 1997.

\bibitem{perelomov86}
A.~Perelomov.
\newblock Generalized coherent states and their applications.
\newblock Springer, Heidelberg, 1986.

\bibitem{kb09}
M.~Ku\'s and I.~Bengtsson.
\newblock {'Classical'} quantum states.
\newblock {\em Phys. Rev. A},{80}:022319, 2009.

\end{thebibliography}

\end{document}